\documentclass[prl,twocolumn,aps,showpacs]{revtex4}
\usepackage{graphicx}% complex graphics
\usepackage{bm}% bold math
\usepackage{amssymb} %math symbols

\begin{document}

\title{Disorder-induced magnetooscillations in bilayer graphene at high bias}

\author{V. V. Mkhitaryan and M. E. Raikh}

\affiliation{ Department of Physics, University of Utah, Salt Lake
City, UT 84112, USA}

\begin{abstract}

Energy spectrum of biased bilayer graphene near the bottom has a
"Mexican-hat"-like shape. For the Fermi level within the Mexican
hat we predict that, apart from conventional magnetooscillations
which vanish with temperature, there are additional
magnetooscillations which are weakly sensitive to temperature.
These oscillations are also insensitive to a long-range disorder.
Their period in magnetic field scales with bias, $V$,  as $V^2$.
The origin of these oscillations is the disorder-induced
scattering between electron-like and hole-like Fermi-surfaces,
specific for Mexican hat.

\end{abstract}
 \maketitle

\begin{figure}[t]
\centerline{\includegraphics[width=90mm,angle=0,clip]{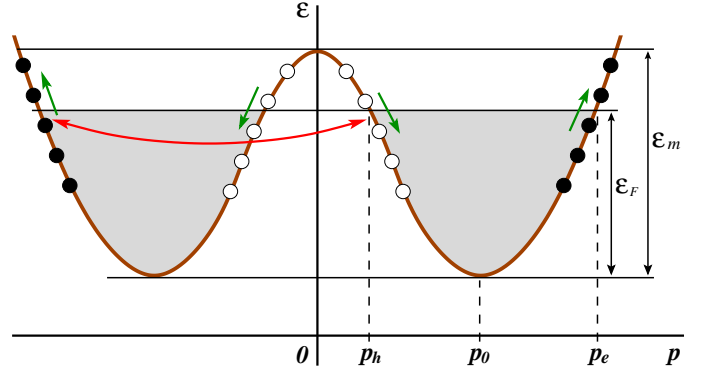}}
\caption{Energy spectrum of a biased bilayer graphene near the
bottom of conduction band. Two Fermi circles, electron-like and
hole-like, have radii $p_e$ and $p_h$, respectively. Black dots
are electron-like Landau levels. White dots are hole-like Landau
levels. Red arrows illustrate scattering process giving rise to
additional oscillations. Green arrows illustrate that energies of
electron-like Landau levels, Eq. (\ref{LLspec}), grow with number,
while energies of hole-like Landau levels fall off with number.}
\label{mexhat}
\end{figure}

{\noindent \it Introduction.} Numerous experimental studies of
electronic properties of bilayer graphene were recently reported
in the literature \cite{Geim06,Savchenko,Geim08,Stormer,
Pinczuk,Ohta, Castro07,Oostinga,Basov08,Basov09,Ujiie,Zettl,
Tutuc, Eisenstein,Kim}. From prospective of potential
applications, the appeal of bilayer graphene  is that a gap can be
opened and tuned by the gate voltage \cite{Ohta,
Castro07,Oostinga,Basov08,Basov09,Ujiie,Zettl, Tutuc,
Eisenstein,Kim}. Using the dual (top and back) gated structures
allows to control both the gap and the carrier density
independently \cite{Oostinga,Zettl,Tutuc,Eisenstein,Kim}. For
these structures, opening of a gap was demonstrated in temperature
\cite{Oostinga} and bias \cite{Oostinga,Zettl} dependence of
resistivity, in bias dependence of capacitance
\cite{Eisenstein,Kim}, as well as in strong-filed magnetotransport
\cite{Tutuc}. Measurements in magnetic field reported in the
literature focused either on weak-field ($B\sim 0.1$~T) domain in
order to reveal weak localization \cite{Savchenko} and universal
conductance fluctuations \cite{Ujiie}, or quantizing ($B\sim
10$~T) fields \cite{Stormer,Ujiie,Castro07,Tutuc, Eisenstein}. At
intermediate fields, $B\sim1$~T, transport and capacitance are
determined by electron states near the band-edge,
Fig.~\ref{mexhat}. As follows from tight-binding calculation by
McCann and Fal'ko \cite{McFal}, the spectrum near the band-edge
has a form of "Mexican hat" with minimum at
\begin{equation}\label{p0}
p_0=\frac{Vt}{v\sqrt{2(V^2+t^2)}},
\end{equation}
where $v=8\times 10^7$ cm/s is the band velocity, $t$ is the
interlayer hoping and $V$ is the bias. The minimum has a depth,
\begin{equation}\label{em}
\varepsilon_m=\frac{V}2\left(1-\frac t{\sqrt{V^2+t^2}}\right).
\end{equation}
The "capacity" of the minimum for $t=0.4$ eV and $V=100$ mV is
$n=p_0^2/(\pi\hbar^2)=0.54\times 10^{12}$ cm$^{-2}$. This density
is comparable to the densities in experiment \cite{Eisenstein},
but the gap in this experiment was as small as $26$ meV. In
experiment \cite{Kim} the gap was wider, $\approx100$ meV, but the
density was higher, $\sim 1.5\times10^{13}$ cm$^{-2}$. In both
experiments, the Fermi energy exceeded $\varepsilon_m$. Upon
increasing the bias from $V=100$ mV to $V=250$ mV, as in
experiment \cite{Zettl}, the capacity increases by a factor $4.8$.
This suggests that in the situation Fig.~\ref{mexhat}, when all
electrons reside in is feasible.

In the present paper we demonstrate that, in the regime
Fig.~\ref{mexhat}, when the Fermi energy is smaller than
$\varepsilon_m$, the behavior of magnetocapacitance, $\delta
C(B)/C^2$, and magnetoresistance, $\delta R(B)/R$, exhibits
oscillations, which are additional to conventional
magnetooscillations related to the Landau levels. They behave as
\begin{equation}\label{addosc}
\frac{\delta C}{C^2}\propto\frac{\delta R}R\propto\cos\bigl[2\pi
p_0^2(V)\lambda^2\bigr],
\end{equation}
where $\lambda=\sqrt{\hbar c/(eB)}$ is the magnetic length and
$p_0$ is the minimum position of the spectrum, given by Eq.
(\ref{p0}). The remarkable feature of the oscillations Eq.
(\ref{addosc}) is that their period does not contain the Fermi
energy, $\varepsilon_{\scriptscriptstyle F}$. This suggests that
they are not smeared upon increasing temperature, while
conventional oscillations are suppressed as
$\exp(-2\pi^2T/\hbar\omega_c)$. For the same reason, additional
oscillations Eq. (\ref{addosc}), unlike conventional
magnetooscillations, are insensitive to the random density
variations caused by long-range disorder. On the other hand,
oscillations Eq. (\ref{addosc}) are {\it disorder-induced}, since
it is scattering by short-range disorder which causes the term Eq.
(\ref{addosc}) in the response functions.
%%%%%%%%%%%%%%%%%%%
%%%%%%%%%%%%%%%%%%

The origin of additional oscillations lies in the fact that, for
$\varepsilon_{\scriptscriptstyle F}<\varepsilon_m$, the Fermi
surface consists of two circles with radii,
$p_e(\varepsilon_{\scriptscriptstyle F})$ and
$p_h(\varepsilon_{\scriptscriptstyle F})$, Fig.~\ref{mexhat}.
Magnetooscillations corresponding to these circles are
$\cos\!\left(\pi p_e^2\lambda^2\right)$ and $\cos\!\left(\pi
p_h^2\lambda^2\right)$, respectively. Due to disorder scattering
between the two Fermi circles, magnetocapacitance and
magnetoresistance will also contain a product,
\begin{eqnarray}\label{oscprod}
&&\cos\!\left( \pi p_e^2\lambda^2\right) \cos\!\left(\pi
p_h^2\lambda^2\right) =\frac12\cos\!\left(\pi
\left[p_e^2-p_h^2\right]\lambda^2\right)\nonumber\\
&&+\frac12\cos\!\left(
\pi\left[p_e^2+p_h^2\right]\lambda^2\right).
\end{eqnarray}
Our prime observation is that $\varepsilon_{\scriptscriptstyle F}$
drops out of the second term. Indeed, the spectrum of bilayer
graphene has a form
\begin{eqnarray}\label{spectrum}
\varepsilon^2(p)= \frac{V^2}4\left(1-2\frac{v^2p^2}{t^2}\right)^2
+\frac{v^4p^4}{t^2},
\end{eqnarray}
so that
\begin{eqnarray}\label{pfermi}
&&p_e^2=p_0^2+ \frac{p_0t}{vV}
\sqrt{2\varepsilon_{\scriptscriptstyle
F}(\varepsilon_{\scriptscriptstyle F}+\sqrt{2}vp_0)},\nonumber\\
&&p_h^2=p_0^2- \frac{p_0t}{vV}
\sqrt{2\varepsilon_{\scriptscriptstyle
F}(\varepsilon_{\scriptscriptstyle F}+\sqrt{2}vp_0)}
\end{eqnarray}
We see that relation $p_e^2+p_h^2=2p_0^2$ holds for arbitrary
$\varepsilon_{\scriptscriptstyle F}<\varepsilon_m$. In the
reminder of the paper we give a derivation of Eq. (\ref{addosc})
with a prefactor and briefly discuss the limit of strong magnetic
fields, where the Hall quantization becomes important.
%Below we trace how the product Eq. (\ref{oscprod}) leading to
%additional oscillations Eq. (\ref{addosc}) in magnetoresistance
%and magnetocapacitance (i.e., density of states at the Fermi
%level) emerges as a result of scattering between electron-like and
%hole-like states.

{\noindent \it Density of states.} We start from zero magnetic
field. Spectrum Eq. (\ref{spectrum}) is a result of
diagonalization of the $2\times2$ matrix Hamiltonian
\begin{eqnarray}\label{ham}
\mathcal{H}=\left(\begin{array}{cc}
\frac V2\left(1-2\frac{v^2p^2}{t^2}\right)&-\frac{v^2(p_x+ip_y)^2}{t}\\
\,\\ -\frac{v^2(p_x-ip_y)^2}{t}&-\frac
V2\left(1-2\frac{v^2p^2}{t^2}\right)
\end{array}\right).
\end{eqnarray}
It is convenient to find the Landau level spectrum of Eq.
(\ref{ham}) using the gauge, ${\bf A}=(0, Bx)$, in which
eigenfunctions of the conventional quadratic spectrum are
$e^{ip_yy}\phi_n(x)$, with $\phi_n(x)$ being the eigenfunctions of
1D harmonic oscillator. In this basis, the spectrum for $n\geq2$
is
\begin{eqnarray}\label{LLspec}
&&\varepsilon_n=-\frac{V}{2t}\hbar\Omega_c\\
&&\pm\sqrt{\!V^2\left(\frac
12-\frac{2n-1}{2t}\hbar\Omega_c\right)^2\!\!
+\left(\hbar\Omega_c\right)^2n(n-1)}\nonumber
\end{eqnarray}
with $\Omega_c=2\hbar v^2/(t\lambda^2)$, while eigenfunctions are
given by
\begin{equation}\label{ef}
\Phi_{n,p_y}({\bf r})= \frac{e^{ip_yy}}{\sqrt{1+d_n^2}}
\!\left(\!\!\begin{array}{c}
\phi_n\left(x-p_y\lambda^2\right)\\
d_n\phi_{n-2}\left(x-p_y\lambda^2\right)
\end{array}\!\!\right),
\end{equation}
where
\begin{equation}\label{ab}
%a_n=\frac1{\sqrt{1+d_n^2}}, \,\,b_n=\frac{d_n}{\sqrt{1+d_n^2}}, \,\,
d_n=\frac{\varepsilon_n-V\left(\frac12-n\frac{\hbar\Omega_c}t\right)}
{\hbar\Omega_c\sqrt{n(n-1)}}.
\end{equation}
Spacings between the electron-like and hole-like Landau levels at
the Fermi energy are the same,
\begin{eqnarray}\label{spac}
\frac{\hbar^2}{\lambda^2}{\bigg |}\frac{\partial
\varepsilon}{\partial p^2}{\bigg |}_{p=p_e}\!\!\!\!\!\!\!&=&
\frac{\hbar^2}{\lambda^2}{\bigg |}\frac{\partial
\varepsilon}{\partial p^2}{\bigg |}_{p=p_h}\\
&=&\frac{\hbar^2}{\lambda^2}\frac{vV}{2p_0t}
\frac{\sqrt{\varepsilon_{\scriptscriptstyle
F}(\varepsilon_{\scriptscriptstyle F}+\sqrt{2}vp_0)}}
{\varepsilon_{\scriptscriptstyle F}+vp_0/\sqrt{2}}=\hbar\omega_c.
\nonumber
\end{eqnarray}

In the presence of disorder, $U({\bf r})$, imaginary part of the
self-energy, $\Sigma_n(E)$, is determined by level numbers, $n$,
for which $\varepsilon_n$ is close to $E$. At the same time, the
$n$-dependence of $\Sigma_n(E)$ is weak. In our case, however, the
equation $\varepsilon_n=E$ has {\it two} different solutions:
$2n_e\simeq p_e^2\lambda^2$ and $2n_h\simeq p_h^2\lambda^2$.
Correspondingly, one should consider two different self-energies,
$\Sigma_{n_e}\equiv\Sigma_e$ and $\Sigma_{n_h}\equiv\Sigma_h$.
%As a result, even in the lowest
%order in disorder strength, in $Im\Sigma$ we have contributions
%from $Im\Sigma_e$ and $Im\Sigma_h$.
As demonstrated in \cite{RaikhShah93}, short-range disorder (with
correlation length smaller than $\lambda$) insures the
applicability of the self-consistent Born approximation (SCBA),
which in our case becomes a system
\begin{eqnarray}
\label{scbae} &&\hspace{-1cm}\frac{\Sigma_e(E)}{\Gamma^2}\! =\!
\sum_{n, e}\! \frac{\alpha_{ee}} {E-\varepsilon_n-\!\Sigma_e(E)}
+\!\sum_{n, h}\!
\frac{\alpha_{eh}} {E-\varepsilon_n-\!\Sigma_h(E)},\\
\label{scbah} &&\hspace{-1cm}\frac{\Sigma_h(E)}{\Gamma^2}\! =\!
\sum_{n,h}\! \frac{\alpha_{hh}} {E-\varepsilon_n-\Sigma_h(E)}
+\!\sum_{n, e}\! \frac{\alpha_{he}} {E-\varepsilon_n-\Sigma_e(E)},
\end{eqnarray}
where subscript $(n,e)$ or $(n,h)$ indicates that the summation is
performed for $n$ close to $n_e$ or $n_h$, respectively.
%Coefficients $a_e$, $a_h$, $b_e$, and $b_h$ do not contain numbers
%of Landau levels since they are taken at $n=n_e$ and $n=n_h$,
%correspondingly.
Coefficients
\begin{equation}\label{alphs}
\alpha_{ee}=\frac{1+|d_{n_e}|^4}{(1+|d_{n_e}|^2)^2},\quad
\alpha_{eh}=\frac{1+|d_{n_e}|^2|d_{n_h}|^2}{(1+|d_{n_e}|^2)(1+|d_{n_h}|^2)}
\end{equation}
do not contain numbers of Landau levels, since they are taken at
$n=n_e$ and $n=n_h$, correspondingly. Explicit expressions for
$d_{n_e}$ and $d_{n_h}$ as a function of the Fermi energy are the
following
\begin{equation}\label{dedh}
d_{n_e}=\frac V t+\frac {t(\varepsilon_{\scriptscriptstyle F}-
\varepsilon_m)} {v^2p_e^2},\,\, d_{n_h}=\frac V t+\frac { t
(\varepsilon_{\scriptscriptstyle F}- \varepsilon_m)}{v^2p_h^2},
\end{equation}
where $\varepsilon_m$ is given by Eq. (\ref{em}), and $p_e$, $p_h$
%as functions of $\varepsilon_{\scriptscriptstyle F}$
are given by Eq.~(\ref{pfermi}). Coefficients $\alpha_{hh}$ and
$\alpha_{he}$ in Eq.~(\ref{scbah}) are given by Eq.~(\ref{alphs})
with replacement, $e\rightleftarrows h$. Coefficient $\Gamma^2$
describes the strength of the disorder potential, $\Gamma^2=\int
d^2{\bf r}\langle U(0)U({\bf r})\rangle/(2\pi\lambda^2)$. Second
terms in Eqs. (\ref{scbae}), (\ref{scbah}) describe contributions
from disorder-induced mixing of electron-like and hole-like Landau
levels.

\begin{figure}[t]
\centerline{\includegraphics[width=90mm,angle=0,clip]{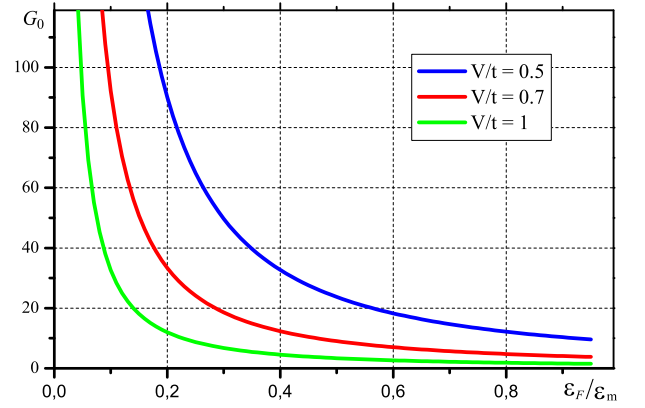}}
\vspace{-0.7cm} \caption{Prefactor in "magnetocapacitance" $\delta
g(B)$, Eq. (\ref{dos}), in the units of $4\pi\Gamma^2\lambda^4
\left[2p_0t/(\hbar^2v V)\right]^3$, is plotted from Eqs.
(\ref{spac}), (\ref{alphs}), and (\ref{dedh}), versus
dimensionless ratio $\varepsilon_{\scriptscriptstyle F}/
\varepsilon_m$ for three values of the ratio, $V/t=0.5$,
$V/t=0.7$, and $V/t=1$.} \label{graf}
\end{figure}

Applying the Poisson summation formula to the sums in Eqs.
(\ref{scbae}), (\ref{scbah}), we get the following equations for
the imaginary parts $\text{Im}\Sigma=\Sigma^{\prime\prime}$,
\begin{eqnarray}
\label{poisse} \Sigma_e^{\prime\prime}
(\varepsilon_{\scriptscriptstyle F})\! &=&\!
\frac{\hbar}{2\tau_e}+\frac{2\pi\Gamma^2}{\hbar\omega_c}
\alpha_{ee}\exp\left[-
\frac{2\pi\Sigma_e^{\prime\prime}}{\hbar\omega_c}\right]
\cos\!\left( \pi p_e^2\lambda^2\right) \nonumber\\
&+&\!\frac{2\pi\Gamma^2}{\hbar\omega_c} \alpha_{eh}\exp\left[-
\frac{2\pi\Sigma_h^{\prime\prime}}{\hbar\omega_c}\right]
\cos\!\left( \pi p_h^2\lambda^2\right),
\\ \label{poissh}
\Sigma_h^{\prime\prime}(\varepsilon_{\scriptscriptstyle F})\! &=&
\! \frac{\hbar}{2\tau_h}+\frac{2\pi\Gamma^2}{\hbar\omega_c}
\alpha_{hh}\exp\left[-
\frac{2\pi\Sigma_h^{\prime\prime}}{\hbar\omega_c}\right]
\cos\!\left( \pi p_h^2\lambda^2\right) \nonumber\\
&+&\!\frac{2\pi\Gamma^2}{\hbar\omega_c} \alpha_{he}\exp\left[-
\frac{2\pi\Sigma_e^{\prime\prime}}{\hbar\omega_c}\right]
\cos\!\left( \pi p_e^2\lambda^2\right),
\end{eqnarray}
where $\tau_e$ and $\tau_h$ are the {\it full} scattering times
from the states $p_e$ and $p_h$ in a zero magnetic field,
\begin{equation}\label{stimes}
\frac{\hbar}{\tau_e}=\frac{2\pi\Gamma^2}{\hbar\omega_c}(\alpha_{ee}
+\alpha_{eh}),\quad
\frac{\hbar}{\tau_h}=\frac{2\pi\Gamma^2}{\hbar\omega_c}(\alpha_{hh}
+\alpha_{he}).
\end{equation}
We note that magnetic-field dependence drops out from the ratio
$\Gamma^2/(\hbar\omega_c)$. Concerning the energy dependence of
$\tau_e$ and $\tau_h$, it follows from Eq. (\ref{spac}) that near
the bottom of Mexican hat, $\varepsilon_{\scriptscriptstyle F}\ll
\varepsilon_m$, we have
$\tau_e,\,\tau_h\sim\sqrt{\varepsilon_{\scriptscriptstyle F}}$,
which reflects the 1D character of the bear density of states
\cite{we}. Iterating Eqs. (\ref{poisse}), (\ref{poissh}), we
obtain a contribution to the density of states of the form
\begin{equation}\label{dos}
\delta g(B)=G_0 \exp\!\!\left[-\frac\pi{\omega_c\tau_e}
-\frac\pi{\omega_c\tau_h}
 \right]\! \cos(2\pi p_0^2\lambda^2),
\end{equation}
which coincides with additional oscillations Eq. (\ref{addosc})
stated in the Introduction. Prefactor $G_0$ is given by
$G_0=4\pi\Gamma^2\alpha_{eh}/\left[(\hbar\omega_c)^3\lambda^2
\right]$. Magnetic field dependence of $G_0$ is $\propto 1/B$.
Energy dependence of $G_0$ is plotted in Fig.~\ref{graf}. We see
that $G_0$ diverges in the limit $\varepsilon_{\scriptscriptstyle
F}\rightarrow 0$. This divergence is also due to the 1D character
of the density of states near the bottom of the Mexican hat.
%Coefficient
%$\alpha_{eh}$ characterizes the efficiency of disorder-induced
%coupling between the electron-like and hole-like states. Its
%evolution with $\varepsilon_{\scriptscriptstyle F}$ as
%$\varepsilon_{\scriptscriptstyle F}$ changes inside the Mexican
%hat is shown in Fig. 2.
%We see that $\alpha_{eh}$ does not change drastically within the
%entire interval $0<\varepsilon_{\scriptscriptstyle
%F}<\varepsilon_m$.
As the Fermi level approaches the top of the Mexican hat, the hole
contributions in Eqs. (\ref{poisse}), (\ref{poissh}), and
resulting additional oscillations, disappear. At the same time the
prefactor $G_0$ remains finite in the limit
$\varepsilon_{\scriptscriptstyle F} \rightarrow \varepsilon_m$.
Such an abrupt behavior of additional oscillations is a
consequence of the fact that the bare density of states
experiences a jump at $\varepsilon_{\scriptscriptstyle F}=
\varepsilon_m$.

{\noindent \it Conductivity} To trace the emergence of the product
$\cos\!\left( \pi p_e^2\lambda^2\right) \cos\!\left( \pi
p_h^2\lambda^2\right)$ in the conductivity, $\sigma$, it is
sufficient to set $\Sigma_e^{\prime\prime}=\hbar/\tau_e$ and
$\Sigma_h^{\prime\prime}=\hbar/\tau_h$ in the exponents in Eqs.
(\ref{poisse}), (\ref{poissh}). The SCBA expression for $\sigma$
in the case of bilayer graphene is the sum of electron-like and
hole-like contributions
\begin{equation}\label{twocont}
\sigma(E)=\sigma_e(E)+\sigma_h(E),
\end{equation}
where $\sigma_e$ can be presented in the form of a sum,
\begin{eqnarray}\label{sige}
&&\sigma_e(E) = \frac{\hbar e^2}{\pi^2
\lambda^2}\\
&&\times\sum_{n,e}\! \frac{\langle v_x \rangle^2_{n,n+1}
\left(\Sigma_e^{\prime\prime}\right)^2} {\left[
(E-\varepsilon_n)^2
+\left(\Sigma_e^{\prime\prime}\right)^2\right]\!\! \left[
(E-\varepsilon_{n+1})^2
+\left(\Sigma_e^{\prime\prime}\right)^2\right]},\nonumber
\end{eqnarray}
and $\sigma_h$ is given by Eq. (\ref{sige}) with replacement of
subindex $e$ by $h$. Matrix element $\langle v_x \rangle_{n,n+1}$
taken between the states Eq. (\ref{ef}) is
\begin{equation}\label{matel}
\langle v_x \rangle_{n,n+1}=\pm i \lambda\omega_c
\frac{\sqrt{n+1}+d_{n}d_{n+1}\sqrt{n-1}}{\sqrt{2(1+|d_{n}|^2)(1+|d_{n+1}|^2)}}.
\end{equation}

Now we notice that a term proportional to the product of two
cosines Eq. (\ref{oscprod}) follows from
$\left(\Sigma_e^{\prime\prime}\right)^2$ in the numerator of Eq.
(\ref{sige}). Indeed, plugging Eq. (\ref{poisse}) into the
numerator of Eq. (\ref{sige}) and replacing
$\left(\Sigma_e^{\prime\prime}\right)^2$ in denominator with its
leading value $\hbar^2/\tau_e^2$, we arrive at additional
oscillations
\begin{equation}\label{cao}
\frac{\delta\sigma_e}{\sigma_e^0}=
\frac{e^2}{2\pi\hbar}\frac{\tau_e^2}{\tau_{ee}\tau_{eh}}
\exp\!\!\left[-\frac\pi{\omega_c\tau_e} -\frac\pi{\omega_c\tau_h}
 \right]\!\cos\!\left(2\pi p_0^2\lambda^2\right),
\end{equation}
where we have introduced the Drude conductivity
\begin{equation}\label{sig0}
\sigma_e^0=\frac{p_e^2\lambda^2}{2\hbar^2}
\frac{\omega_c\tau_e}{1+\omega_c^2\tau_e^2}.
\end{equation}
A more accurate form of $\delta\sigma_e/\sigma_e^0$ can be found
by applying the Poisson summation formula to Eq. (\ref{sige}),
which will include corrections to self-energies in the
denominator. This transforms Eq. (\ref{cao}) into
\begin{eqnarray}\label{acao}
\frac{\delta\sigma_e}{\sigma_e^0}&=&
\frac{e^2}{2\pi\hbar}\frac{\tau_e^2}{\tau_{ee}\tau_{eh}}
\frac{1-\omega_c^2\tau_e^2}{1+\omega_c^2\tau_e^2}\\
&\times&\exp\!\!\left[-\frac\pi{\omega_c\tau_e}
-\frac\pi{\omega_c\tau_h}
 \right]\!\cos\!\left(2\pi p_0^2\lambda^2\right),\nonumber
\end{eqnarray}
which differs from Eq. (\ref{cao}) in strong, $\omega_c\tau_e>1$,
magnetic fields. For the contribution $\sigma_h$, relations Eq.
(\ref{acao}) holds with replacement of subindexes
$e\rightleftarrows h$.
%Incorporating of magnetic field in the
%semiclassical limit reduces to discretizing the momenta,
%$p_n=\sqrt{2n}/\lambda$ \cite{LifKos}.
%%%%%%%%%%%%%%%%%%%
%%%%%%%%%%%%%%%%%%

{\noindent \it Concluding remarks.} ({\it i}) Up to now, observing
the Mexican hat structure of the spectrum in experiments on
bilayer graphene was limited by relatively low mobility,
$\mu\sim2000-3000$ cm$^2$/Vs. This corresponds to the energy
smearing, $\hbar/\tau$, of about $7-10$ meV. In particular,
revealing Landau quantization in magnetotransport
\cite{Geim06,Castro07,Ujiie,Tutuc}, ac \cite{Stormer}, and
magnetocapacitance \cite{Eisenstein, Kim} experiments required
strong magnetic field, $B\sim 10$ T. Typical cyclotron quantum for
such fields is $\hbar\omega_c\sim 20$~meV, i.e., it is bigger than
$\varepsilon_m\approx  11$ meV for $V=0.2$ V . On the other hand,
inhomogeneity of local electron density \cite{Eisenstein} was a
significant factor in smearing of magnetooscillations. In this
regard, additional oscillations Eq.~(\ref{addosc}), being
insensitive to this inhomogeneity, might be observable even when
conventional magnetooscillations are completely washed out. For
$B=1$ T and same $V=0.2$~V condition $\varepsilon_m>\hbar\omega_c$
is satisfied; for this $V$, the product $p_0^2\lambda^2$ in the
argument of Eq. (\ref{addosc}) is $38$. For such fields,
conventional oscillations are suppressed at temperatures as low as
$T=3$~K, while additional oscillations Eq. (\ref{addosc}) remain
unaffected. Unlike conventional magnetooscillations \cite{NCD},
they are also insensitive to the lifting of valley degeneracy.

({\it ii}). Our calculation was based on the spectrum
Eq.~(\ref{spectrum}); this spectrum is obtained from $2\times2$
Hamiltonian Eq.~(\ref{ham}). Analysis of more general $4\times4$
Hamiltonian \cite{McFal,Review09} suggests that the gap can exceed
$t$ while the property, $p_e^2+p_h^2=2p_0^2$ persists.

({\it iii}). Relevant densities for the additional oscillations
are $\sim 10^{12}$ cm$^{-2}$. Such densities are high enough for
electron-electron interaction-induced spectrum renormalization to
be insignificant \cite{FermiRing,MacDonald,Viola}. On the other
hand, interactions can scatter electrons between electron-like and
hole-like Fermi surfaces. They also induce inelastic lifetime
$\sim\varepsilon_{\scriptscriptstyle F}/T^2$. This leads to
effective suppression of additional oscillations at temperatures
above $T\sim \sqrt{\varepsilon_{\scriptscriptstyle
F}\hbar\omega_c}\sim 50$ K .

({\it iv}). To establish a relation between oscillation
Eq.~(\ref{addosc}) and magnetointersubband oscillations in a
quantum well with two subbands \cite{RaikhShah94}, let us turn to
the product Eq.~(\ref{oscprod}) of the oscillating part of the
density of states. Magnetointersubband oscillations of Ref.
\cite{RaikhShah94} follow from the similar product for different
subbands. However, they emerge from the term,
$\cos[\pi(p_e^2-p_h^2)\lambda^2]$, of Eq. (\ref{oscprod}), while
oscillations Eq. (\ref{addosc}) come from the term
$\cos[\pi(p_e^2+p_h^2)\lambda^2]$ of Eq. (\ref{oscprod}).
Independence of this term of $\varepsilon_{\scriptscriptstyle F}$
is specific for bilayer graphene.
%In quantum well two groups of
%levels belong to different subbands, while in our case they are
%electron-like and hole-like levels from the same subband.

({\it v}). In closing, we discuss qualitatively the limit of
quantizing magnetic fields. When the Fermi level lies within the
Mexican hat, classical trajectories corresponding to electron-like
and hole-like states are Larmour circles with {\it opposite}
direction of rotation. Indeed, the equation of motion in momentum
space, $\dot{\bf p}=\frac ec\frac{\partial\varepsilon}{\partial
{\bf p}}\times {\bf B}$, can be presented as
\begin{equation}\label{clasem}
\dot{\bf p}=\frac ec\left(\frac{vV}{p_0\,t}\right)^2
\frac{p^2-p_0^2}{2\varepsilon} \left({\bf p}\times{\bf B}\right).
\end{equation}
With energy and absolute value of momentum conserved by Eq.
(\ref{clasem}), the only difference between electron-like and
hole-like motions comes from the factor, $(p^2-p_0^2)$. Since
$(p^2_e-p_0^2)=-(p^2_h-p_0^2)$, clockwise rotation of
electrons-like states and anti-clockwise rotation of hole-like
states have the same frequency, in agreement with Eq.
(\ref{spac}). At the same time, the radii and velocities of their
Larmour motions are related as $p_e/p_h$.

Opposite directions of rotation for electron-like and hole-like
states translate into the opposite sings of drift velocities for
corresponding edge states,
\begin{eqnarray}\label{vevh}
&&v_e=\left(\frac{vV}{p_0\,t}\right)^2
\frac{p_e^2-p_0^2}{2\varepsilon}\frac{\sqrt{p_e^2-p_y^2}}
{\pi-\arccos(p_y/p_e)},\nonumber\\
&&v_h=\left(\frac{vV}{p_0\,t}\right)^2
\frac{p_h^2-p_0^2}{2\varepsilon}\frac{\sqrt{p_h^2-p_y^2}}
{\pi-\arccos(p_y/p_h)}.
\end{eqnarray}
This, in turn, means that dispersion laws for electron-like and
hole-like states {\it intersect} each other. Previously, Refs.
\cite{Levitov,Fertig} pointed out that opposite dispersion of the
edge states from the {\it same} Landau level can arise from the
valley splitting. Combined with the Zeeman splitting, this leads
to intersecting edge dispersions for opposite spin directions
\cite{Levitov}. We note that in bilayer graphene with the Fermi
level within the Mexican hat, crossing of the edge dispersions
from {\it different} Landau levels occurs naturally with the
valley degeneracy preserved. For the Fermi level located at the
intersection of electron- and hole-dispersion curves, interactions
can result in non-chiral Luttinger liquid at the edge. This
situation is similar to the quantum Hall line junction considered
in Refs. \cite{Fradkin1, Fradkin2}. Unlike Refs. \cite{Fradkin1,
Fradkin2}, where the disorder results in resonant-tunelling states
between the edges separated by a tunnel barrier, in our case
disorder will smear the corresponding Luttinger-liquid anomalies.

\end{document}